# Security and Reliability Evaluation of Countermeasures implemented using High-Level Synthesis

Amalia-Artemis Koufopoulou [(1)], Kalliopi Xevgeni [(1)], Athanasios Papadimitriou [(1) (2)], Mihalis Psarakis [(1)] and David Hely [(3)]

[(1)] Dept. of Informatics, University of Piraeus, Piraeus, Greece
[(2)] Dept. of Digital Systems, University of the Peloponnese, Greece
[(3)] Univ. Grenoble Alpes, Grenoble INP, LCIS, Valence, France

*Abstract*— As the complexity of digital circuits increases, High-Level Synthesis (HLS) is becoming a valuable tool to increase productivity and design reuse by utilizing relevant Electronic Design Automation (EDA) flows, either for Application-Specific Integrated Circuits (ASIC) or for Field Programmable Gate Arrays (FPGA). Side Channel Analysis (SCA) and Fault Injection (FI) attacks are powerful hardware attacks, capable of greatly weakening the theoretical security levels of secure implementations. Furthermore, critical applications demand high levels of reliability including fault tolerance. The lack of security and reliability driven optimizations in HLS tools makes it necessary for the HLS-based designs to validate that the properties of the algorithm and the countermeasures have not been compromised due to the HLS flow. In this work, we provide results on the resilience evaluation of HLS-based FPGA implementations for the aforementioned threats. As a test case, we use multiple versions of an on-the-fly SBOX algorithm integrating different countermeasures (hiding and masking), written in C and implemented using Vivado HLS. We perform extensive evaluations for all the designs and their optimization scenarios. The results provide evidence of issues arising due to HLS optimizations on the security and the reliability of cryptographic implementations. Furthermore, the results put HLS algorithms to the test of designing secure accelerators and can lead to improving them towards the goal of increasing productivity in the domain of secure and reliable cryptographic implementations.

## I. INTRODUCTION

High-Level Synthesis (HLS) tools can effectively enhance the productivity for the design of complex digital circuits by automatically creating Register Transfer Level (RTL) descriptions given high-level algorithms as input [1]. This productivity enhancement stems from the fact that the designers do not need to worry about the details of the comparatively cumbersome and potentially error-prone RTL design phase. HLS tools are nowadays capable of efficiently synthesizing the high-level code in an optimized fashion for key constraints, such as power consumption, area and timing [2]. Furthermore, it is very efficient for the designers to perform a design space exploration so as to obtain an RTL design that fits the demands of their project [3]. Additionally, such tools require considerably less effort for the design to be verified and the verification can start very early in the design flow [4].

The advantages of HLS can though become a major drawback when traditional constraints are not the only ones of interest [5]. Furthermore, when key design objectives cannot be described at the high level of abstraction that HLS tools accept as inputs, it can potentially be a recipe for failure [6]. To make things worse, the designers may not be able to easily monitor such issues at the generated RTL due to its complexity [7]. For all the aforementioned reasons, HLS designers need to be careful that HLS optimizations do not compromise quality.

Nowadays, digital circuits used in safety-critical applications require high levels of security [8]. One of the strengths of using HLS to design secure circuits is that it is easy to integrate security functions such as crypto accelerators [9] to more complex circuits. When secure hardware components are integrated into digital circuits, implemented either as ASICs or FPGAs, each with its own development considerations, hardware security and reliability are major concerns [10,11].

Electronic devices are susceptible to naturally occurring faults from their operating environment which may affect their reliability [12]. Furthermore, faults may be deliberately induced by adversaries to extract secret information [13,14]. Side Channel Analysis (SCA) attacks are another category of hardware attacks which may reduce or even cancel the security levels of secure hardware components [15]. In order to protect against such threats, cryptographic algorithms need to employ appropriate countermeasures and mitigation techniques depending on the necessary level of resilience.

Most countermeasures tackling those issues utilize techniques that may be introduced in several different abstraction levels of the design flow depending on their architecture (High-Level Language (HLL), RTL, Gate-level, physical design). Considering an HLS design flow, in order to maintain the productivity, scalability and design reuse, the protection techniques need to be integrated at the highest level of abstraction [16]. Nevertheless, if the HLS tool neglects to address optimization goals compatible with secure and reliable circuits, then such a design flow may become not only inefficient but potentially even dangerous [17,18]. Therefore, for HLS tools to be part of a productive design flow for secure and reliable hardware design, they need to optimize such protection techniques in order to guarantee security and reliability levels [18].

In this work, our goal is to evaluate the impact of an HLS-based design flow on the security and reliability of unprotected and protected cryptographic functions implemented using modern FPGA devices against SCA and Fault Injection (FI) attacks. To this end, we have implemented an evaluation platform capable of performing SCA and FI evaluations on HLS-generated RTL cryptographic circuits. To the best of our knowledge, so far, only a few works try to evaluate the capability of HLS tools to maintain security properties of countermeasures protecting against SCA attacks, as detailed in the Related Work. Furthermore, most works on SCA focus their analysis on the effects of HLS optimizations to the memory elements storing the result of an algorithm, such as the AES SBOX output [19, 20, 21]. Concerning the impact of faults on the behavior of countermeasures added at the HLL of an HLS flow for cryptographic implementations, some early steps have been made towards hardening the generated circuits [22, 23]. In this work, we analyze the results of FI campaigns on countermeasures synthesized using an HLS flow. This way we show results on the effects of various HLS optimizations on the effectiveness and resiliency of the countermeasures.

Our goal is to provide evidence of issues arising due to HLS optimizations on the security and the reliability of cryptographic implementations. We believe that such an analysis can be useful to HLS designers who are not necessarily hardware security experts or low-level hardware designers so as to avoid dangerous effects of HLS on secure implementations. Furthermore, the presented results put HLS algorithms to the test of designing secure accelerators and can lead to improving them towards the goal of increasing productivity in the domain of secure and reliable cryptographic implementations.

## II. Related Work

To protect cryptographic implementations from SCA attacks, designers resort to the integration of countermeasures, which include numerous protection techniques [15, 18], but are mainly categorized as hiding- or masking-based [24]. Regarding security against such threats when HLS is used, literature presents a limited number of works. In [18], the fact that the modern HLS tools disregard security concerns is highlighted, but at the same time, it is pointed out that their workflow can assist in the integration of secure mechanisms. In [25], by comparing a manually developed RTL design with an HLS-generated one, regarding an FPGA crypto core, the authors show that HLS can be efficiently used when the target architecture is taken into account in the developing stage. A similar view and results are presented in [21]. In addition, the concept of resource diversification and parallelism is presented, as a way to limit SCA leakage. This falls into the category of hiding countermeasures. In [26], the authors propose a Boolean masking scheme that can be used from HLS. All their designs focus on Look-Up-Table (LUT)-based implementations, and therefore, their SCA evaluations concern the memory elements in which the outputs of the cryptographic algorithms are stored. Apart from that, they explain that their masking scheme does not apply well in cryptographic mechanisms of higher complexity due to the presence of glitches. In [19] the authors compare SCA resistance of various components developed either at RTL or using an HLS flow. They execute SCA evaluations to multiple LUT-based implementations of the AES SBOX. Specifically, they perform the attacks to the output of the SBOX, and thus their HLS optimizations under investigation are limited to the storage elements used for the SBOX outputs. Even though the authors claim that some HLS implementations are less vulnerable, they still leak secret information. The same authors had previously examined in [20] the security of different memory elements, by performing HLS resource pragma exploration. They provide similar results while focusing again on the SCA leakage of the output of the SBOX.

Fault analysis attacks have become a prominent subject in literature [11,13,14], with new attacks constantly emerging [27]. In any case, the effect of (even unintended) soft errors that could tamper with the functionality of a design should always be a consideration for hardware developers, especially when cryptographic applications are of interest [28,29]. To improve the fault-tolerance, multiple mitigation techniques and countermeasures have been proposed [30,31,32]. It is crucial thus, to understand and evaluate the behavior of a design from early development stages. For that reason, simulation or emulation-based evaluations at the RTL are important [33].

Reliable HLS methodologies, such as the one proposed in [34], introduce the concept of reliability as a design metric, and present a theoretical approach to quantify it. Integration of such a metric in HLS could allow for the design exploration of reliable applications, with respect to area and latency limitations. In [35], the authors extend this idea by introducing an automated approach for the synthesis of reliable designs based on genetic algorithms. In [36], the authors suggest that the core operations of HLS flow, scheduling and binding, could be modified in order to satisfy a given threshold of reliability. In [37], the lifetime of variables within the circuit is examined, which appears to be related to their vulnerability. The aim is to minimize it through reliability-aware HLS operations.

Few works evaluate their results with the use of FI campaigns. For instance, the authors in [22] work inside the HLS flow to incorporate an encoding scheme for arithmetic operations, whereas in [23], the authors aim to insert redundancy from C-code variable specifications. To the best of our knowledge, there are no related works that study the effects of HLS optimizations on the robustness of protected or unprotected cryptographic implementations under the malicious or unindented injection or the inherent occurrence of faults.

## III. Design Under Evaluation and Countermeasures

We have used the Canright SBOX algorithm [38] as a well-known security test case. Two countermeasures against SCA attacks have been integrated in the high-level algorithm: a) a hiding-based scheme that also supports fault detection and b) a masking-based protection scheme. Furthermore, the unprotected version and the two protected versions of the Canright algorithm have been synthesized through the Xilinx Vivado HLS flow using three different sets of directives to test the effects of commonly used synthesis settings to reach typical design optimization goals. To evaluate all these designs in terms of performance, security and reliability, we have fully implemented them in FPGA and developed two experimental evaluation flows: the first one performs SCA attacks on the Canright implementations in an FPGA board and the second performs FI campaigns on the RTL designs produced by the HLS tool. As the Canright algorithm computes the SBOX of a given input on-the-fly, it was necessary to evaluate the security of the entire algorithm by considering, besides the final output of the SBOX (which is usually considered), all the intermediate values of the Canright algorithm. Therefore, our analysis evaluates the effects of the HLS to the entire SBOX, including all internal functions and their optimizations.

Even though more optimized SBOX implementations exist in the literature [39], we chose to use Canright's SBOX. This is because our goal was to show the effects of HLS on countermeasures and thus, in our view, it was more appropriate to demonstrate our methodology on a well-established implementation than the state-of-the-art solution. Furthermore, we decided to focus our analysis on the SBOX module instead of a complete cryptographic algorithm (such as AES) in order to simplify the complexity of the algorithm and enhance the understanding of the issues arising due to HLS optimizations.

The design is capable of performing the SBOX computations on-the-fly, through the use of tower-field representation. Instead of performing the $7^{th}$-degree polynomial inversion required for SBOX, the data are transformed to be represented in isomorphic, lower-degree fields, where the inversion is easier. Specifically, the inversion of a $1^{st}$-degree polynomial (a two-bits value) corresponds to a bit swap. The countermeasures presented below were developed over this concept.

In the following subsections, we shortly describe the countermeasures that we have added to the SBOX algorithm, named hereafter as "Unprotected HLS" implementation or "UHLS" in abbreviation.

## A. Correlated Noise Generation (CNG)

This countermeasure is based on the addition of correlated noise to the computation of the SBOX by duplicating it and using the same data and a fake key in parallel to the true secret key computation [40]. The fact that the data are the same between the two parallel computations helps to hide the leakage of the true secret key [41]. The CNG countermeasure parallelizes the computation of the two bytes by concatenating them into a single 16-bit variable. The datapath of the SBOX has been extended by changing the data type of the algorithm's variables from uint8_t to int, as shown in Fig. 1. This way the least significant byte is used for the computations of the true key and the second byte for the fake key. This CNG technique has been chosen as a typical hiding-based SCA countermeasure, but since it uses redundancy it can also be considered as a duplication-based fault detection countermeasure.

## B. Masking

Masking countermeasures attempt to de-correlate the power traces and the secret key by using random values to mask the data. When the computations are completed the masks are removed from the SBOX output and the correct results are obtained. Our masking scheme has been previously presented in [42] and adopts the tower-field approach. An 8-bit input mask is randomly generated and XOR-ed with the input data. Three additional random masks are also generated as shown in Fig. 2. Along with two consecutively transformed values of the input mask, -(input mask)' and output mask-, they are able to conceal the intermediate SBOX results. Their size corresponds to the representation field they are applied to – two 8-bit masks for $GF(2^8)$, two 4-bit masks for $GF(2^4)$, and one 2-bit mask for $GF(2^2)$. The result is generated carrying the output mask, and can be removed by applying the output mask with an XOR operation. It is known that this masking countermeasure is vulnerable in the presence of glitches and other physical defaults [43]. This is useful for our analysis since it can show the degree to which specific HLS optimizations will lead to RTL designs (later on synthesized by low-level FPGA tools) with different amounts of glitches and physical defaults that may weaken the countermeasure.

## C. HLS Optimizations & Directives

In order to put under test the HLS optimization effects on the security of our countermeasures, we have selected three meaningful sets of directives supported by the Vivado HLS tool, leading to three solutions (Sol) for each design. For all the design-solution combinations, we have set the timing constraints to 20ns. Sol1 is the default optimization strategy and serves as a reference. It concerns minimal optimizations, such as small functions' inlining, small loops' unrolling and use of registers for small arrays' elements. Sol2 performs a full loop unrolling to the design and targets the lowest possible latency.

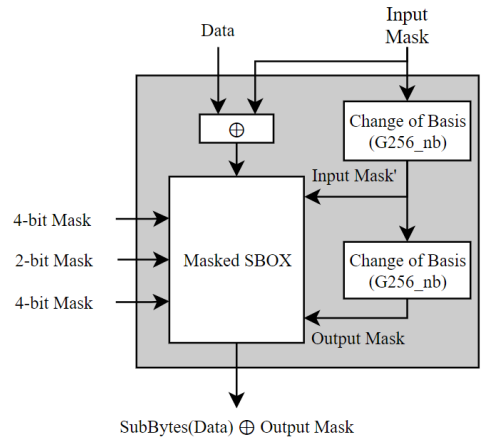

Figure 2. Masked SBOX architecture

Sol3 turns inlining off for all the Canright SBOX functions, hence preserving all function sub-modules without flattening them. Furthermore, it enables the use of BRAM resources to store the change of basis arrays as well as the output of the function responsible for the change of basis (G256_nb).

## D. FPGA resources utilization

Table I presents the programmable resource utilization for each design and solution. By "UHLS" we denote the original open-source C-version Canright SBOX and by "Verilog" the original Verilog version, while the remaining designs are the protected versions. "Design Resources" column presents the utilized programmable resources in terms of BRAMs, FFs and LUTs as well as the maximum latency (in clock cycles) for the synthesis of each design and solution. "Overheads vs UHLS" column presents the improvement in resource utilization with respect to the UHLS solutions. It should be noted that the utilization statistics, except for the Verilog implementation, were derived from the Vivado HLS synthesized designs, hence they are estimates of the actual statistics.

Sol1 involves the largest amount of design resources among all cases, while Sol2 greatly reduces the FFs and increases the LUTs due to loop unrolling. As expected, it produces designs with the lowest latency scores. Sol3 achieves lower resource utilization concerning FFs and LUTs than Sol1. The use of BRAMs is also noted. This optimization strategy achieves latency scores slightly increased compared to Sol1. Lastly, the manually designed Verilog implementation achieves a latency score of one clock cycle and has by far the lowest resource utilization of all HLS implementations studied in our work.

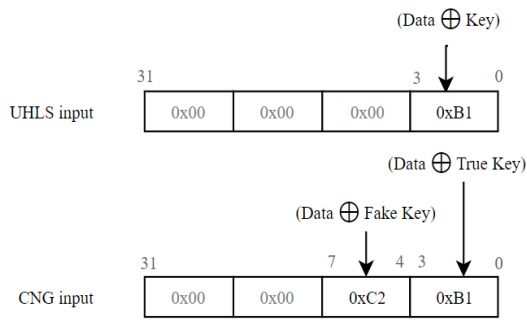

Figure 1. UHLS (top) and CNG architecture (bottom)

Table I. Programmable resource utilization estimates and overheads compared to UHLS designs (Latency: in clock cycles)

|  |  | Design Resources | | | Overheads vs UHLS | | |
|---|---|---|---|---|---|---|---|
|  |  | Sol1 | Sol2 | Sol3 | Sol1 | Sol2 | Sol3 |
| UHLS | BRAM_18K | 0 | 0 | 3 | 0 | 0 | 0 |
|  | FF | 293 | 18 | 109 | 0 | 0 | 0 |
|  | LUT | 599 | 446 | 446 | 0 | 0 | 0 |
|  | Latency | 34 | 1 | 37 | 0 | 0 | 0 |
| CNG | BRAM_18K | 0 | 0 | 3 | 0 | 0 | 0 |
|  | FF | 212 | 50 | 207 | -81 | 32 | 98 |
|  | LUT | 1607 | 1930 | 1386 | 1008 | 1484 | 940 |
|  | Latency | 36 | 3 | 38 | 2 | 2 | 1 |
| Masked | BRAM_18K | 0 | 0 | 3 | 0 | 0 | 3 |
|  | FF | 583 | 169 | 373 | 293 | 18 | 109 |
|  | LUT | 1206 | 1312 | 1226 | 599 | 446 | 446 |
|  | Latency | 59 | 4 | 59 | 0 | 0 | 0 |
| Verilog | BRAM_18K | 0 | | | - | | |
|  | FF | 111 | | | - | | |
|  | LUT | 152 | | | - | | |
|  | Latency | 1 | | | - | | |

## IV. EVALUATION FLOW

### A. SCA Experimental Setup

We used the ChipWhisperer CW305 FPGA evaluation board employing an Artix7 Xilinx FPGA (XC7A100T-2FTG256) to evaluate our designs against SCA. To acquire the power traces, we used the ChipWhisperer differential amplifier for power analysis and an oscilloscope involving a maximum sampling rate of 2GSPS and 350MHz bandwidth. To perform the data acquisition, we have developed a Matlab software, which provides the input to the SBOX residing in the FPGA and records the results for further processing. At the same time, the acquisition software collects for each SBOX computation the power traces through the oscilloscope.

### B. SCA Evaluation Flow

Our evaluation platform performs SCA on all the intermediate operations (C-code operations) of a given C algorithm as well as the Welch t-test. This way, it can connect the leakage of a Correlation Power Analysis (CPA) attack to the intermediate operation and analyse the entire computation. The evaluation flow is based on the standard CPA using Pearson's formula to compute the correlation coefficients between the power traces and the intermediate results of the SBOX algorithm for all key hypotheses under a power model. In our case, the software supports both Hamming Weight (HW) and Hamming Distance (HD) power models. Additionally, we have developed a tool capable of producing the intermediate values for which we will perform the CPA attacks. Instead of performing the attacks only against the output of the SBOX algorithm, our tool reads the HLS C code and creates one internal intermediate value for each SBOX operation. For the Canright SBOX algorithm, our tool generated 338 different internal intermediate values. Therefore, instead of performing a single CPA attack, we perform 338 attacks for each power model (HW and HD) and each design and solution under evaluation. Therefore, our attack model considers that the attacker has concluded that the Canright SBOX is a candidate for the implementation under attack and uses the open-source C code to generate any possible internal intermediate value.

By attacking all the operations of the design under evaluation, we can monitor the effects of the HLS optimizations for each design in greater detail. Thus, we can also use this higher resolution to evaluate internal optimizations of the SBOX instead of only memory-related optimizations already performed in previous approaches as described in the Related Work. Furthermore, this approach can directly measure the side-channel leakage of each operation. Additionally, we apply to all designs an evaluation based on Welch's t-test [44], which examines whether two sets of data, consisting of the traces derived for different inputs, can be distinguished. If so, it means that the leakage is strongly related to the input, and also to the cryptographic key. The degree of distinguishability is given by the value $t$, calculated for each set's mean $\mu$, variance $s^2$ and number of elements $n$:

$$t = \frac{\mu_0 - \mu_1}{\sqrt{\frac{s_0^2}{n_0} + \frac{s_1^2}{n_1}}}$$

Two sets of data were used, one consisting of 100,000 traces generated for random data inputs, and one consisting of 100,000 traces generated for a constant input. It is assumed that a t-value $|t|$ greater than 4.5 indicates a side channel leakage capable of exposing the secret key.

### C. Fault Injection Evaluation Flow

We have implemented a FI platform based on the Xilinx Vivado tool that uses commands of the Vivado simulator. The potential fault sites of our FI campaigns are the Flip-Flops of the elaborated RTL netlist generated by HLS synthesis without any further optimizations. Our fault model assumes bit flip in one flip-flop (Single Bit-Flip, SBF) or bit flips in more than one flip-flop (Multiple Bit-Flip, MBF) for a specific clock cycle. Our FI campaigns include exhaustive SBF experiments (i.e., faults in all the Flip-Flops and all clock cycles) as well as statistical MBF experiments for different multiplicities. In order to generate the fault sites (flip-flop) and the injection times (clock cycle), we use the statistical approach proposed in [33].

## V. RESULTS

### A. SCA Evaluation Results

Table II presents the results of the SCA in all the protected and unprotected designs. The first column presents all the target designs including the Verilog implementation and all the combinations of the HLS optimization (Sol1, Sol2, Sol3) and the SBOX scheme (UHLS, CNG and Masked). The designs are in this order to facilitate the discussion on the results. UHLS is the unprotected HLS implementation; CNG is the HLS-based version of the CNG countermeasure; Masked design is the HLS-based masked SBOX. Sol1, Sol2 an Sol3 are the three HLS optimization settings described in Section III.C. The values shown in the table depict the number of successful attacks, i.e., attacks that expose the secret cryptographic key.

For each evaluation, we have used 100,000 power traces. Therefore, each table cell shows the number of internal SBOX operations (intermediate values) that expose the secret key using the HW or the HD power model (HW and HD results are added). Column "Total" provides the total number of SBOX operations leaking the secret key, while the other columns categorize them according to which function of the Canright SBOX the corresponding operations belong to. The Verilog implementation leaks the secret key only in 7 different operations, which shows that it is much more secure against SCA attacks than the unprotected HLS implementation and most protected designs. Concerning Sol1, it is apparent that the unprotected HLS (UHLS) is the weakest design involving 51 leaking operations. Then, we observe that by adding the countermeasures while using Sol1, we can gradually achieve higher protection reducing the leaking operations to 19 for the CNG and 9 for the Masked design. Regarding Sol2, the results are not so straightforward as the UHLS design leaks the key in less operations (32 in total) than the protected designs (58 for the CNG and 56 for the Masked design). As the main characteristic of this optimization directive (Sol2) is the loop unrolling, we can see that the heavier the optimization is, the worst is the security of the countermeasures. Therefore, heavy HLS optimizations, which are unsupervised by the designer, may lead to a less secure design than the unprotected implementation, even if a countermeasure is included. Sol3 further increases the number of leaking operations for all the designs besides the Masked one. This fact can be attributed to the use of BRAM since it is the only major difference with the default optimization option (Sol1). By examining the RTL, we determined that the default inlining performed in Sol1 has no effect in the Masked scheme, since it would only concern simple functions. From the other Sol3 designs, "UHLS" reaches 87 leaky operations. Similar to Sol1, in Sol3 we can also observe that the addition of countermeasures improves the security of the algorithm. CNG reduces the leaky operations to 28, while the Masked design achieves the lowest number of successful attacks amongst all designs and optimization cases.

*Table II. Number of successful attacks (HW and HD) in different SBOX operations per design.*

|  |  | Total | Sbox | G256_nb | G256_inv | G16_sq_scl | G16_mul | G16_inv | G4_scl_N | G4_scl_N2 | G4_mul | G4_sq |
|---|---|---|---|---|---|---|---|---|---|---|---|---|
|  | Verilog | 7 | 2 | 0 | 0 | 0 | 0 | 0 | 3 | 0 | 0 | 2 |
| sol1 | UHLS | 51 | 4 | 31 | 4 | 0 | 1 | 0 | 3 | 0 | 8 | 0 |
| sol1 | CNG | 19 | 0 | 10 | 2 | 0 | 1 | 1 | 0 | 0 | 5 | 0 |
| sol1 | Masked | 9 | 1 | 1 | 1 | 0 | 0 | 2 | 0 | 0 | 4 | 0 |
| sol2 | UHLS | 32 | 1 | 18 | 4 | 0 | 9 | 0 | 0 | 0 | 0 | 0 |
| sol2 | CNG | 58 | 0 | 0 | 2 | 0 | 15 | 10 | 4 | 0 | 27 | 0 |
| sol2 | Masked | 56 | 0 | 0 | 3 | 0 | 29 | 7 | 0 | 0 | 15 | 2 |
| sol3 | UHLS | 87 | 6 | 43 | 10 | 0 | 20 | 0 | 6 | 0 | 2 | 0 |
| sol3 | CNG | 28 | 0 | 11 | 3 | 0 | 2 | 1 | 0 | 0 | 11 | 0 |
| sol3 | Masked | 2 | 1 | 1 | 0 | 0 | 0 | 0 | 0 | 0 | 0 | 0 |

A significant result is that even powerful countermeasures can be greatly impacted by HLS optimizations. The Masked design highlights greatly this point. It is important to emphasize that the masked SBOX theoretically should be 100% protected against first-order CPA attacks. Our results show that the Masked design contains multiple operations which leak the secret key even against a first-order CPA attack. Furthermore, we can observe that the Masked countermeasure works better with the BRAM optimization (Sol3), where it leaks the key only in 2 operations. The most strong hypothesis for this result is that when using the BRAM for the masked design, there is a reduction in glitches existing in the operation of this implementation. As described in [44], glitches are a big concern for masked implementations. Therefore, for such designs, it is essential for HLS designers to be careful with the HLS optimizations they use so as not to increase the likelihood of generating glitches in the design. In Fig.3, we show the evaluation results for the Masked-Sol3 implementation. The tool generated 338 figures including the correlation at each sampling point for all key hypotheses. Using the tools, we can determine that the two operations that leak the secret key belong to the G256_inv and G16_inv functions.

In Fig. 4, Welch's t-test verifies multiple CPA results discussed above. Sol3 is vulnerable for all designs while the most secure design is the Masked one. On the other hand, for t-test the unprotected UHLS and the CNG designs are all vulnerable for all optimization scenarios. This is also due to the fact that the CPA is performed on the unprotected SBOX intermediate values and that CNG is a hiding-based countermeasure.

### B. Fault Injection Results

Based on the FI experiments, we assess the reliability and the security of the selected designs. Since we have used a single SBOX, all fault injections which lead to capturing errors at the outputs are candidates for successful fault attacks (e.g.

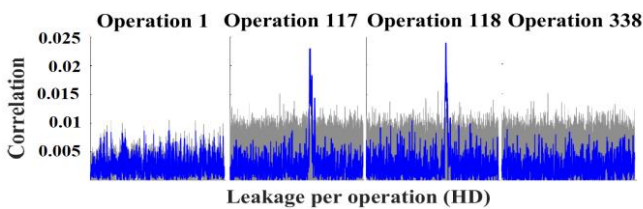

*Figure 3. Evaluation results on the Masked- Sol3 design*

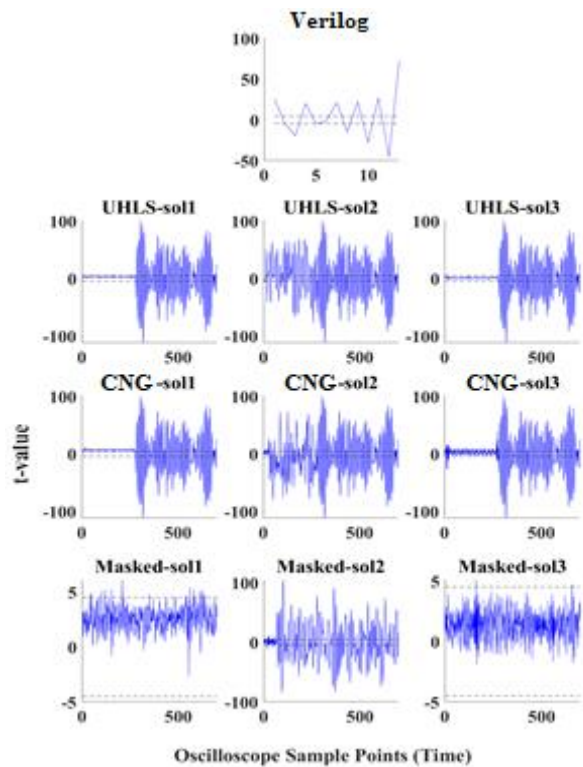

*Figure 4. 2-point (Welch's) t-test results*

by using single-bit or single-byte Differential Fault Analysis) [11,13]. Concerning reliability, all errors at the outputs of the SBOX may completely change the entire AES output depending on the round that the fault be injected. Fig. 5 presents the results of the FI campaigns. Our fault mode includes exhaustive SBFs and statistical MBFs with multiplicities of 2, 3, 4 and 5 (denoted as m2, m3, m4 and m5, respectively). We have selected the number of samples for each experiment so as to achieve a margin of error of 1% with a confidence level of 99%. In the cases where the SBOX is unprotected against FI attacks (e.g., masked and UHLS), each fault is characterized as Silent (i.e., no errors are captured at the SBOX output), Critical (i.e., leads to an erroneous output) and Hang (i.e., does not allow the completion of the computation). For the CNG, which integrates a fault detection scheme, we further characterize the faults as detected (i.e., the fault results in different outputs for the two redundant computations).

The first column presents the results for the designs synthesized using the default HLS optimization (Sol1) goals. UHLS involves the higher critical error rates between the three designs; they range from 4.36% for SBFs to 16.16% for MBFs-m5. The results for the CNG countermeasure show a drastic reduction of critical error rates compared to UHLS and Masked. We should note that CNG manages to detect all SBFs leading to critical errors. Additionally, we notice a gradual increase in the corresponding detection rates as the fault multiplicities increase. On the other hand, this design involves the largest hang rates among all designs for the Sol1 case. The Masked implementation involves critical errors ranging between 1.81% for SBFs to 7.66% for MBFs-m5. The second column presents the results for the optimization solutions involving complete loop unrolling (Sol2). For the UHLS design, the critical faults increase drastically to 50% for the SBF campaign while they reach 68.46% for larger multiplicities. This drastic deterioration is due to the fact that complete loop unrolling results to a circuit including just 18 flip flops and a la-tency of one clock cycle. When CNG is implemented by enforcing complete loop unrolling, critical error rates vary from 0% for SBFs to 0.78% for MBFs-m5. Once again we can see that CNG manages to detect all SBFs leading to critical errors. Similarly Masked SBOX yields smaller critical error rates than UHLS. The larger amount of silent errors in the cases of CNG and Masked designs is due to the higher number of flip flops and latency than the unprotected design. The third column presents the results for Sol3. The main property of this strategy is that unlike Sol1 it does not allow inlining of the design's functions. The results for UHLS show that in comparison to Sol1 and Sol3, it has slightly higher critical error rates. The CNG design, when synthesized without inlining (Sol3), yields the lowest critical error rates, 0% for SBFs up to 0.14% for MBFs-m5. This is expected as not allowing to perform optimizations across functions reduces resource sharing. The Masked implementation for Sol3 shows less critical error rates compared to UHLS and the other Masked solutions.

For all the CNG solutions, we can notice that the critical rates increase with the fault multiplicity. For a perfectly duplicated design, we would expect that SBFs would result in one of the two redundant modules to fail, and thus, it would theoretically yield very high rates of detection. This means that the HLS implementations of CNG avoid resource sharing and manage to keep the two redundant computations unaffected for all solutions. The Masked implementation is the most complex design. This translates to a difficulty of the HLS tool to optimize the benefits of inlining towards greater resource sharing. Thus, for all the Masked versions, the less efficient inlining contributes to less resource sharing leading to a tendency for less critical error rates than UHLS. Globally, we remark that HLS optimizations affect the resilience of either protected or unprotected designs against SCA or FI attacks.

## VI. CONCLUSIONS

In the current work, we have studied the impact of using an HLS flow on the security of protected and unprotected cryptographic implementations against SCA attacks. We have implemented the Canright SBOX and two protected versions of the same algorithm integrating hiding and masking countermeasures using Vivado HLS. We have used three different HLS optimization strategies to achieve various design goals. Additionally, we have evaluated all the designs and optimization strategies in terms of security and reliability performing SCA and FI experiments. Our results highlight the fact that secure circuit designers should be careful when using an HLS flow to integrate SCA countermeasures. Globally, we remark that HLS optimizations affect the resilience of either protected or unprotected designs against SCA or FI attacks. An important result is that the HLS tools manage to maintain the properties of the implemented countermeasures to some extent. Concerning SCA, the theoretically stronger masked implementation achieves the best evaluation results among the tested designs besides when loop unrolling took place. Our results also show that, even though masking achieves higher protection, it breaks after 100,000 traces. Similarly, under FI evaluations, the duplication-based CNG countermeasure maintains its theoretical ability to detect all single bit flips for all tested constraints. Our results show that designers can exploit the HLS tools to greatly improve productivity at the cost of extra care that has to be taken when security and reliability are important goals. Additionally, we show that there is a need for further research and development to enhance HLS algorithms in order to transparently take into account the need for secure and reliable hardware accelerators.

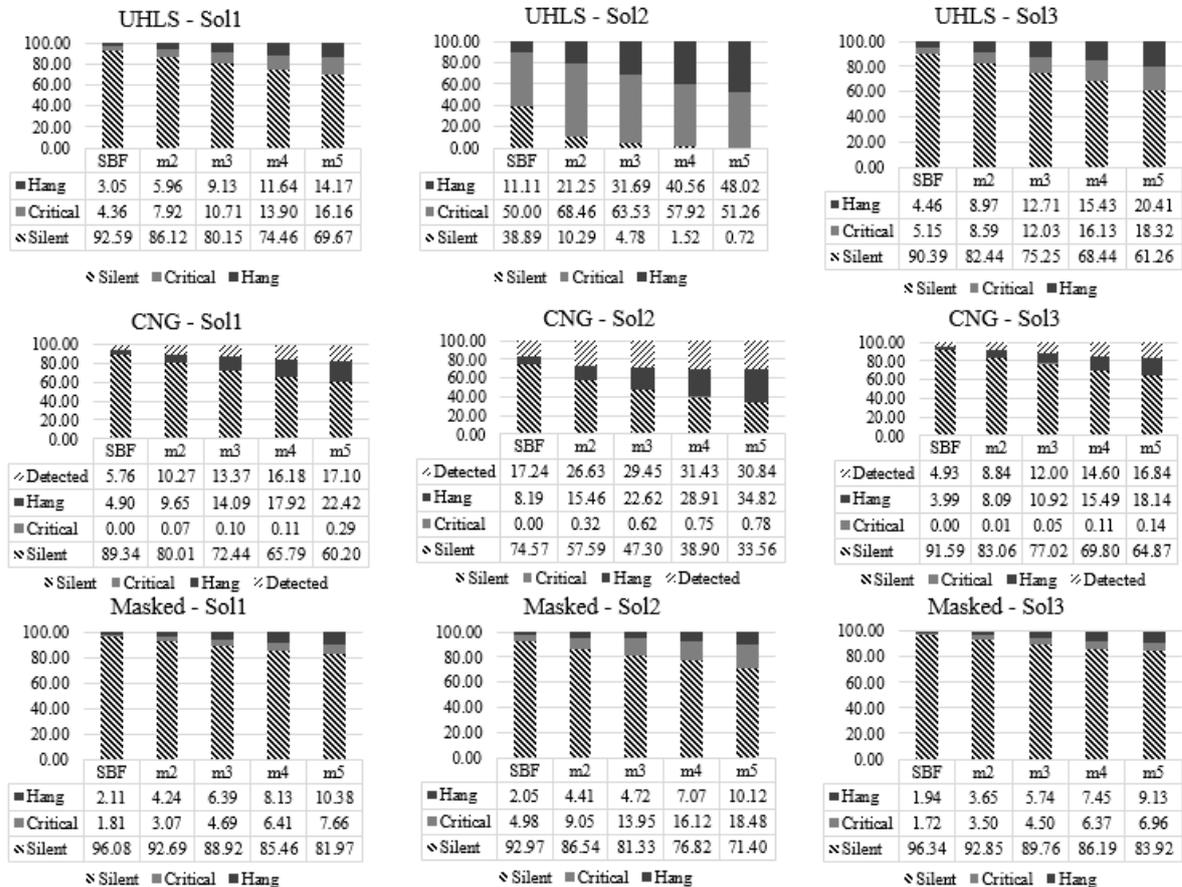

*Figure 5. Fault injection effects over the designs under examination*

## REFERENCES

[1] Coussy, Philippe, et al. "An introduction to high-level synthesis." *IEEE Design & Test of Computers* 26.4 (2009): 8-17.

[2] Nane, Razvan, et al. "A survey and evaluation of FPGA high-level synthesis tools." *IEEE Transactions on Computer-Aided Design of Integrated Circuits and Systems* 35.10 (2015): 1591-1604.

[3] Choi, Young-kyu, and Jason Cong. "HLS-based optimization and design space exploration for applications with variable loop bounds." *2018 IEEE/ACM International Conference on Computer-Aided Design (ICCAD)*. IEEE, 2018.

[4] Takach, Andres. "Design and verification using high-level synthesis." *2016 21st Asia and South Pacific Design Automation Conference (ASP-DAC)*. IEEE, 2016.

[5] Zhang, Lu, et al. "Machine-learning-based side-channel leakage detection in electronic system-level synthesis." *IEEE Network* 34.3 (2020): 44-49.

[6] Elliott, John P. *Understanding behavioral synthesis: a practical guide to high-level design*. Springer Science & Business Media, 1999.

[7] Bailey, Donald G. "The advantages and limitations of high level synthesis for FPGA based image processing." *Proceedings of the 9th International Conference on Distributed Smart Cameras*. 2015.

[8] Acosta, Antonio J., Tommaso Addabbo, and Erica Tena-Sánchez. "Embedded electronic circuits for cryptography, hardware security and true random number generation: an overview." *International Journal of Circuit Theory and Applications* 45.2 (2017): 145-169.

[9] Chang, Jed Kao-Tung, et al. "Hardware-assisted security mechanism: The acceleration of cryptographic operations with low hardware cost." *International Performance Computing and Communications Conference*. IEEE, 2010.

[10] Mukhopadhyay, Debdeep, and Rajat Subhra Chakraborty. *Hardware security: design, threats, and safeguards*. CRC Press, 2014.

[11] Joye, Marc, and Michael Tunstall, eds. *Fault analysis in cryptography*. Vol. 147. Heidelberg: Springer, 2012.

[12] Kastensmidt, Fernanda Lima, Luigi Carro, and Ricardo Augusto da Luz Reis. *Fault-tolerance techniques for SRAM-based FPGAs*. Vol. 1. Dordrecht: Springer, 2006.

[13] Biham, Eli, and Adi Shamir. "Differential fault analysis of secret key cryptosystems." *Annual international cryptology conference*. Springer, Berlin, Heidelberg, 1997.

[14] Bar-El, Hagai, et al. "The sorcerer's apprentice guide to fault attacks." *Proceedings of the IEEE* 94.2 (2006): 370-382..

[15] Thillard, Adrian, Emmanuel Prouff, and Thomas Roche. "Success through confidence: Evaluating the effectiveness of a side-channel attack." *International Conference on Cryptographic Hardware and Embedded Systems*. Springer, Berlin, Heidelberg, 2013.

[16] Sadhukhan, Rajat, Sayandeep Saha, and Debdeep Mukhopadhyay. "Shortest Path to Secured Hardware: Domain Oriented Masking with High-Level-Synthesis." *2021 58th ACM/IEEE Design Automation Conference (DAC)*. IEEE, 2021.

[17] K. Tiri, "Side-Channel Attack Pitfalls," *2007 44th ACM/IEEE Design Automation Conference*, 2007, pp. 15-20.

[18] Pilato, Christian, Siddharth Garg, Kaijie Wu, Ramesh Karri, and Francesco Regazzoni. "Securing hardware accelerators: A new challenge for high-level synthesis." *IEEE Embedded Systems Letters* 10, no. 3 (2017): 77-80.

[19] Zhang, Lu, Dejun Mu, Wei Hu, Yu Tai, Jeremy Blackstone, and Ryan Kastner. "Memory-based high-level synthesis optimizations security exploration on the power side-channel." *IEEE Transactions on Computer-Aided Design of Integrated Circuits and Systems* 39, no. 10 (2019): 2124-2137.

[20] Zhang, Lu, Wei Hu, Armaiti Ardeshiricham, Yu Tai, Jeremy Blackstone, Dejun Mu, and Ryan Kastner. "Examining the consequences of high-level synthesis optimizations on power side-channel." In *2018 Design, Automation & Test in Europe Conference & Exhibition (DATE)*, pp. 1167-1170. IEEE, 2018.

[21] Silitonga, Arthur, et al. "Hls-based performance and resource optimization of cryptographic modules." *2018 IEEE Intl Conf on Parallel & Distributed Processing with Applications, Ubiquitous Computing & Communications, Big Data & Cloud Computing, Social Computing & Networking, Sustainable Computing & Communications (ISPA/IUCC/BDCloud/SocialCom/SustainCom)*. IEEE, 2018.

[22] Campbell, Keith A., et al. "High-level synthesis of error detecting cores through low-cost modulo-3 shadow datapaths." *2015 52nd ACM/EDAC/IEEE Design Automation Conference (DAC)*. IEEE, 2015.

[23] Lojda, Jakub, et al. "Data types and operations modifications: A practical approach to fault tolerance in HLS." *2017 IEEE East-West Design & Test Symposium (EWDTS)*. IEEE, 2017.

[24] Mangard, Stefan, Elisabeth Oswald, and Thomas Popp. *Power analysis attacks: Revealing the secrets of smart cards*. Vol. 31. Springer Science & Business Media, 2008.

[25] Homsirikamol, Ekawat, and Kris Gaj. "Can high-level synthesis compete against a hand-written code in the cryptographic domain? A case study." In *2014 International Conference on ReConFigurable Computing and FPGAs (ReConFig14)*, pp. 1-8. IEEE, 2014.

[26] Socha, Petr, Vojtěch Miškovský, and Martin Novotný. "High-level synthesis, cryptography, and side-channel countermeasures: A comprehensive evaluation." *Microprocessors and Microsystems* 85 (2021): 104311.

[27] Dobraunig, Christoph, et al. "SIFA: exploiting ineffective fault inductions on symmetric cryptography." *IACR Transactions on Cryptographic Hardware and Embedded Systems* (2018): 547-572.

[28] Bandeira, Vitor, et al. "Impact of radiation-induced soft error on embedded cryptography algorithms." *Microelectronics Reliability* 126 (2021): 114349.

[29] Sheikhpour, Saeide, Ali Mahani, and Nasour Bagheri. "Practical fault resilient hardware implementations of AES." *IET Circuits, Devices & Systems* 13.5 (2019): 596-606.

[30] Karaklajić, Duško, Jörn-Marc Schmidt, and Ingrid Verbauwhede. "Hardware designer's guide to fault attacks." *IEEE Transactions on Very Large Scale Integration (VLSI) Systems* 21.12 (2013): 2295-2306.

[31] Maistri, Paolo, et al. "Countermeasures against EM analysis for a secured FPGA-based AES implementation." *International Conference on Reconfigurable Computing and FPGAs (ReConFig)*. IEEE, 2013.

[32] Baksi, Anubhab, et al. "A Survey On Fault Attacks On Symmetric Key Cryptosystems." *ACM Computing Surveys (CSUR)* (2022).

[33] Leveugle, Régis, et al. "Statistical fault injection: Quantified error and confidence." *2009 Design, Automation & Test in Europe Conference & Exhibition*. IEEE, 2009.

[34] Tosun, Suleyman, et al. "Reliability-centric high-level synthesis." *Design, Automation and Test in Europe*. IEEE, 2005.

[35] Taher, Farah Naz, Mostafa Kishani, and Benjamin Carrion Schafer. "Design and optimization of reliable hardware accelerators: Leveraging the advantages of high-level synthesis." *2018 IEEE 24th International Symposium on On-Line Testing And Robust System Design (IOLTS)*. IEEE, 2018.

[36] Shastri, Aniruddha, Greg Stitt, and Eduardo Riccio. "A scheduling and binding heuristic for high-level synthesis of fault-tolerant FPGA applications." *IEEE 26th International Conference on Application-specific Systems, Architectures and Processors (ASAP)*. IEEE, 2015

[37] Chen, Liang, Mojtaba Ebrahimi, and Mehdi B. Tahoori. "Reliability-aware resource allocation and binding in high-level synthesis." *ACM Transactions on Design Automation of Electronic Systems (TODAES)* 21.2 (2016): 1-27.

[38] Canright, David. "A very compact S-box for AES." *International Workshop on Cryptographic Hardware and Embedded Systems*. Springer, Berlin, Heidelberg, 2005.

[39] Reyhani-Masoleh, Arash, Mostafa Taha, and Doaa Ashmawy. "Smashing the implementation records of AES S-box." *IACR Transactions on Cryptographic Hardware and Embedded Systems* (2018): 298-336.

[40] Kamoun, Najeh, Lilian Bossuet, and Adel Ghazel. "Correlated power noise generator as a low cost DPA countermeasures to secure hardware AES cipher." *2009 3rd International Conference on Signals, Circuits and Systems (SCS)*. IEEE, 2009.

[41] Aerabi, Ehsan, Athanasios Papadimitriou, and David Hely. "On a side channel and fault attack concurrent countermeasure methodology for MCU-based byte-sliced cipher implementations." *2019 IEEE 25th International Symposium on On-Line Testing and Robust System Design (IOLTS)*. IEEE, 2019.

[42] Canright, David, and Lejla Batina. "A very compact "perfectly masked" S-box for AES." *International Conference on Applied Cryptography and Network Security*. Springer, Berlin, 2008.

[43] Schneider, Tobias, and Amir Moradi. "Leakage assessment methodology." *International Workshop on Cryptographic Hardware and Embedded Systems*. Springer, Berlin, Heidelberg, 2015.

[44] Barthe, Gilles, et al. "maskverif: Automated verification of higher-order masking in presence of physical defaults." *European Symposium on Research in Computer Security*. Springer, Cham, 2019.